\documentstyle[preprint,tighten,prc,aps,epsfig]{revtex}

\begin{document}

\title{Incoherent Eta Photoproduction from the Deuteron near Threshold}

\author{A. Sibirtsev$^1$, S. Schneider$^1$, Ch. Elster$^{1,2}$, 
J. Haidenbauer$^1$, S. Krewald$^1$ and J. Speth$^1$}

\address{
$^1$Institut f\"ur Kernphysik, Forschungszentrum J\"ulich,
D-52425 J\"ulich \\
$^2$Institute of Nuclear and Particle Physics,
Ohio University, Athens, OH 45701}
\date{\today}

\maketitle

\begin{abstract}
Very recent data for the reaction  $\gamma{d}{\to}\eta{np}$,  
namely total cross sections,
angular and momentum spectra, are 
analyzed within a model
that includes contributions from the impulse approximation  and 
next order corrections due to the $np$ and $\eta{N}$ interactions
in the final state. Comparison between the calculations and the
new data indicate sizable contributions from the $np$ and $\eta{N}$
final state interactions.
Some systematic discrepancies between
the calculations and the data are also found.

\end{abstract}

\pacs{}
\maketitle

The reaction  $\gamma{d}{\to}\eta{np}$ close to the meson production 
threshold  offers an opportunity 
to investigate the final state interactions (FSI) between the outgoing 
particles, proton, neutron and $\eta$-meson. Provided the FSI between the 
nucleons is understood, the reaction allows to draw conclusions about the
$\eta$N interaction at low energies.
In Refs.~\cite{Our1,Our2} we
investigated incoherent $\eta$ photoproduction from the deuteron close to
threshold taking into account that 
the reaction amplitude is given by the sum of the first order term,
the impulse 
approximation (IA), and the terms of next higher order due to
the final state
interactions in the neutron-proton ($np$)  
the $\eta$-nucleon ($\eta{N}$) system.

When comparing our calculations to the data~\cite{Krusche1} available at 
that time we found that the few  experimental points for the
cross section of the reaction
$\gamma{d}{\to}\eta{np}$ close to the
reaction threshold require for an adequate description
the additional  contribution from the $\eta$-nucleon
final state interaction.


In Fig.\ref{maeta10} these old data points are indicated by open squares.
It seemed clear that more precise measurements are necessary to 
further understand the interplay between the final state 
interactions of the two subsystems. Moreover, in \cite{Our1} we concluded 
that due to the strong $np$ final state interaction the 
momentum spectra of the $\eta$-meson should be enhanced at high 
momenta if the reaction  is considered close to threshold.

Very recently the TAPS Collaboration reported new data~\cite{Hejny}
for the  incoherent photoproduction of $\eta$-mesons  from the deuteron near
threshold. These new data not only contain 
the total   $\gamma{d}{\to}\eta{np}$ reaction cross section but also
angular and momentum spectra of the $\eta$-meson. In this brief communication
we  study whether those 
new data~\cite{Hejny} can shed some light on issues raised in our 
previous work~\cite{Our1,Our2}.

For the sake of completeness 
we briefly summarize the main ingredients of our previous 
calculations \cite{Our2} on
incoherent photoproduction of $\eta$-mesons from the deuteron.  
Let us recall  that for a given spin $S$ and isospin $T$ 
of the final nucleons the amplitude ${\cal M}_{IA}$ for the 
impulse approximation is written as
\begin{equation}
{\cal M}_{IA}{=}A^T(s_1)\phi(p_2){-}(-1)^{S+T}A^T(s_2)
\phi (p_1),
\end{equation}
where  $\phi(p_i)$ is the deuteron wave function~\cite{Machleidt1}, 
$p_i (i=1,2)$ is the
momentum of the proton or neutron in the deuteron rest frame, and $A^T$ is
the isoscalar or isovector $\eta$-meson photoproduction amplitude 
on a nucleon at the squared invariant collision energy $s_N$
\begin{equation}
s_N=s-m_N^2-2(E_\gamma+m_d)E_N+2\vec{k}_\gamma\cdot\vec{p}_i.
\end{equation}
The  photon momentum is given by ${k}_\gamma$, and the invariant mass by
$s{=}m_d^2{+}2m_dE_\gamma$,
 $E_N$ is the total nucleon energy
and $m_N$ and $m_d$ are the nucleon and deuteron mass, respectively. 

Now  the amplitude ${\cal M}_{NN}$ for the $NN$ final state 
interaction is given by
\begin{equation}
{\cal M}_{NN}=m_N\int dk\, k^2\, \frac{t_{NN}(q,k)\, A^T(s_N^\prime) \,
\phi(p_N^\prime)}{q^2-k^2+i\epsilon}, 
\end{equation}
where $q$ is the nucleon momentum in the final $np$ system and
\begin{equation}
\vec{p}_N^{\, \prime}=\vec{k}+\frac{\vec{k}_\gamma -\vec{p}_\eta}{2},
\end{equation}
$p_\eta$ is the $\eta$-meson momentum and 
$p_N^\prime{=}|\vec{p}_N^{\,\prime}|$.
The half-shell $np$ scattering matrix $t_{NN}(q,k)$ in the
$^1S_0$ and $^3S_1$ partial waves  was obtained
at corresponding off-shell momenta $k$
from the CD-Bonn potential~\cite{Machleidt1}.

Finally, the amplitude ${\cal M}_{\eta N}$ for the $\eta$N 
final state interaction is given as

\begin{equation}
{\cal M}_{\eta N} {=} \frac{m_N m_\eta}{m_N+m_\eta}\!
\int \! dk\,  k^2  \, \frac{t_{\eta N}(q,k) \,
A^T(s_N^{\prime\prime})\,   \phi (p_N^{\prime\prime})}
{q^2-k^2+i\epsilon},
\label{etanfsi}
\end{equation}
where the $\eta$-meson momenta in the final and intermediate 
state of the $\eta{N}$ system are indicated by $q$ and $k$,
$t_{\eta N}(q,k)$ is the half-shell $\eta{N}$  scattering 
matrix in the $S_{11}$ partial wave and
\begin{equation}
\vec{p}_N^{\,\prime\prime}= \vec{k} +
\frac{m_N \, (\vec{k}_\gamma-\vec{p}_N)}{m_N+m_\eta},
\end{equation}
where $\vec{p}_N$ is the momentum of final proton or neutron in the
deuteron rest frame and $m_\eta$ is $\eta$-meson mass.

Furthermore, within the effective range approximation, 
the $\eta{N}$  on-shell scattering matrix is related to the 
scattering length $a_{\eta N}$ as
\begin{equation}
\left[ iq{-}\frac{1}{ a_{\eta N}}\right]^{-1}\!\!\!
{=}\pi \frac{\sqrt{q^2+m_N^2}\sqrt{q^2+m_\eta^2}}
{\sqrt{q^2+m_N^2}+ \sqrt{q^2+m_\eta^2}}\,\, t_{\eta N}(q,q).
\label{efra}
\end{equation}

In our previous work \cite{Our1,Our2} we showed that within
our approach  the uncertainty of the calculations is
dominated by the  insufficient knowledge of the strength of the $\eta$N interaction at
low energies, here represented by $a_{\eta N}$. Moreover, possible
effects due to higher order corrections from the multiple scattering 
expansion~\cite{Delborgo,Gillespie} might be overshadowed by the
sizable variation
of $a_{\eta N}$, which as a result of different model calculations or
extractions can range from 0.25+i0.16 
to 1.05+i0.27~fm. In our calculations we adopt $t_{\eta N}(q,k)$ from
the J\"ulich meson-baryon model~\cite{Krehl}, which gives a scattering length
$a_{\eta N}$=0.42+i0.32~fm. 
When comparing our calculations~\cite{Our1} with the old cross section data 
for the reaction $\gamma{d}{\to}\eta{np}$~\cite{Krusche1}, we concluded that
the value of $a_{\eta N}$ given by this model was consistent with the data.


A comparison between our full calculation, including NN and $\eta$N FSI,
 with the recent experimental information~\cite{Hejny} 
for the  total cross section of the reaction  $\gamma{d}{\to}\eta{np}$ is shown in 
Fig.~\ref{maeta10}. Here the new data are indicated by filled
circles. We can well describe the data close to the reaction
threshold, while there is systematic underprediction of $\simeq$10\%
of the experimental results between 660 and 680~MeV photon energy. We should 
not attribute this discrepancy to $a_{\eta N}$, since 
we found in Ref.~\cite{Our2} that the 
$\eta{N}$ interaction acts predominantly  very close to threshold.
We also want to point out that our calculation matches up with 
the older data (open squares) at energies larger than 680~MeV.


The recent data of Ref.~\cite{Hejny} are more complete and contain not only
total cross sections but also angular distributions 
of $\eta$-mesons in the photon-deuteron
center-of-mass (c.m.) system at different photon energies. They 
are shown in Fig.~\ref{mainz5a}
together with our calculations. 
The angular spectra  at the lower energies, 630${\le}E_\gamma{\le}$650~MeV, 
are quite sensitive to both final state interactions. Especially, the $\eta N$ 
FSI is necessary to describe the data. 

At photon energies 650${\le}E_\gamma{\le}$689~MeV our predictions
show a stronger peaking at forward angles compared to the data,  
and a slight but 
systematic underestimation of the data at backward angles. 
The latter might be attributed to an additional contribution from 
re-scattering mechanism with intermediate $\pi$-meson and 
$\pi{N}{\to}\eta{N}$ transition. However, we are aware that we can not make any
final assessment about the discrepancies at the present stage.

The momentum spectra of the $\eta$-mesons in the $\gamma{-}d$ cm frame
are shown in Fig.~\ref{mainz5b} and compared with our calculations.
We would like to emphasize that the theoretical results displayed represent 
an average over a finite energy interval. This is done in order to 
make the predictions comparable to the experiment, where likewise an 
averaging over energy bins is made \cite{Hejny}.
Specifically for the momentum distribution   of the
$\eta$ meson this averaging has a significant influence on the result as
shown in \cite{Our1}.
The vertical arrows  in Fig.~\ref{mainz5b}  indicate the  maximally allowed  
$\eta$ momentum, which is  given by
\begin{equation}
p_\eta=\frac{[(s{-}(m_n{+}m_p)^2{-}m_\eta^2)^2{-}
4(m_n{+}m_p)^2m_\eta^2]^{1/2}}
{2\sqrt{s}},
\label{max}
\end{equation}
and calculated for the maximal photon energy $E_\gamma$ indicated
in the figure.


As Fig.~\ref{mainz5b} clearly indicates,  for the
lowest energy interval a substantial part of 
experimental points  is located beyond the kinematical limit. 
This might~\cite{Private} stems from a larger experimental 
uncertainty in determining the $\eta$-meson momentum. These 
errors  are
not indicated in  Fig.~\ref{mainz5b} by  horizontal error bars.
Unfortunately, due to this a clean  comparison between
our calculations and the data can not be made.  As an aside, when shifting 
all data points by the same percentage inside the kinematically
allowed region, the  cross section points fall
closer toward our calculation. However this can only serve as 
a guide to the eye and does not 
allow any further speculations.

In order to investigate the sensitivity of the $\eta$ momentum distribution
to the
$\eta{N}$ scattering length we calculate $\eta$-meson angular 
and momentum spectra at  $E_\gamma$=630-640 and 640-650~MeV
with different values for  $a_{\eta N}$ and compare our results with
the data~\cite{Hejny} in Fig.~\ref{mainz5u}. Here the solid lines show
our calculations with $a_{\eta N}$=0.42+i0.32~fm, the dashed lines
the ones with 
$a_{\eta N}$=0.74+i0.27~fm, and the dotted line the calculations with 
$a_{\eta N}$=0.25+i0.16~fm. As it turns out, both observables are quite
sensitive to the size of $a_{\eta N}$.

It can be instructive to  consider different possibilities of 
analyzing the data 
to find a representation which may shed a different light on the
reaction  $\gamma{d}{\to}\eta{np}$. For this reason we consider the 
Dalitz plot representation, which is given as
\begin{equation}
\frac{d\sigma}{ds_{\eta p}ds_{np}}= \frac{|{\cal M}_{IA}{+}{\cal M}_{NN}
{+}{\cal M}_{\eta N}|^2}{256\pi^3 s(s-m_d^2)}.
\end{equation}
Here $s_{\eta p}$ and $s_{np}$ denote the squared invariant mass of the
$\eta{p}$ and $np$ subsystems. 

This representation may be an additional tool to isolate the different FSI. 
In Fig.~\ref{invar1} we display the  Dalitz plot projections calculated at
photon energies $E_\gamma$=643 and 681~MeV. For this calculation we
employ the $\eta N$ FSI given by the J\"ulich model with 
$a_{\eta N}$=0.42+i0.32~fm. The hatched areas indicate 
the contributions from the impulse approximation, the dotted  lines stand
for the  the calculations 
with $np$ FSI alone, while the solid lines represent the full calculations.
The difference between the impulse approximation  and full calculation is
not only a result of the absolute sizes,  but essentially 
the different shapes of the invariant mass spectra. At
$E_\gamma$=643~MeV the low mass part of $np$ spectrum is
substantially enhanced by the $np$ FSI. Thus the
$\eta{p}$ spectrum is shifted to higher masses. 
The difference between the calculations with $\eta{N}$
and $np$ FSI and that with $np$ alone can be considered as
an overall rescaling of the model results. This can be
well understood through our findings in Ref.~\cite{Our2},
namely 
that a quite weak $\eta{N}$ interaction can manifest itself
through the interference with the substantially stronger
$np$ FSI.

The results become more exciting at $E_\gamma$=681~MeV.
While the shape of the $\eta{p}$ distribution is almost 
similar to that obtained at $E_\gamma$=643~MeV, the $np$ 
spectrum clearly shows a low mass structure due to the 
$np$ and $\eta{N}$ FSI, and the size of this
enhancement is given by the coherent sum of the $np$ and $\eta{N}$ 
interactions. The production mechanism alone,  or the contribution
of the impulse approximation may well be 
isolated by imposing $np$ invariant mass cuts. 


We believe that an experimental observation of 
such double peaks structure might serve as  direct
evidence of FSI effects. Finally, we notice that the Dalitz plot 
analysis of the reaction $pp{\to}pp\eta$ measured at COSY~\cite{TOF} 
indicates quite a similar structure in the $pp$ invariant mass
distribution. This finding may encourage further
analysis of the new  $\gamma{d}{\to}\eta{np}$ data~\cite{Hejny}.

In conclusion, we presented  a detailed comparison between our model
for the reaction $\gamma{d}{\to}\eta{np}$ 
developed in Refs.~\cite{Our1,Our2} and recently published experimental 
information~\cite{Hejny} on total cross sections as well as
angular and momentum $\eta$-meson spectra for this reaction.
For our calculations we employ
the $\eta{N}$ FSI obtained  from the J\"ulich
meson-baryon model. 

The comparison between the data~\cite{Hejny} and our calculations 
shows reasonably agreement. The $\gamma{d}{\to}\eta{np}$ data close
to the reaction threshold require additional contributions from  the
$\eta{N}$ FSI and are consistent with  the size of $a_{\eta N}$=0.42+i0.32~fm
given by the J\"ulich model. The angular and momentum $\eta$-meson 
spectra at $E_\gamma{\le}$650~MeV are very sensitive to the size
of $a_{\eta N}$.

However, we found some $\simeq$10\% disagreement between
our calculations and the new data~\cite{Hejny} for the total cross section
for the reaction 
$\gamma{d}{\to}\eta{np}$  at  photon energies 
650${\le}E_\gamma{\le}$689~MeV.  Furthermore, at these
energies the model predicts a stronger peaking of the angular
distribution at forward angles  
and a slight but systematic  underestimation of the data at 
backward angles. Further investigations are
necessary in order to clarify whether this discrepancy stems 
from re-scattering mechanisms. 

In addition, we found that the Dalitz plot analysis of the reaction
$\gamma{d}{\to}\eta{np}$ may serve as 
a very helpful tool for isolating FSI effects. The
Dalitz plot projection on $np$ invariant mass may show a
clean double peak structure at $E_\gamma$=681~MeV,
while at low photon energy $E_\gamma$=643~MeV the
$np$ invariant mass spectrum is substantially enhanced
at low masses.

\begin{acknowledgments}
This work was performed in part under the auspices of the 
U.~S. Department of Energy under contract No. DE-FG02-93ER40756 
with the Ohio University.  The authors appreciate valuable 
discussions with V.~Hejny, B.~Krusche, V.~Metag and H.~Str\"oher,
and thank the TAPS Collaboration for providing us 
with the new experimental results.
\end{acknowledgments}


\newpage

\begin{figure}
\caption{ The total cross section for inclusive photoproduction of $\eta$
mesons off deuterium as function of the photon energy $E_{\gamma}$.
The open squares are old data~\protect\cite{Krusche1}, while the 
circles indicates new results~\protect\cite{Hejny}. The dotted line 
represents the IA calculation, while the dashed line is the 
result with the $np$ final state interaction. The solid line shows
the full calculation, including the $\eta N$ final state interaction 
from the J\"ulich meson-baryon model~\protect\cite{Our2}. } 
\label{maeta10}
\end{figure}

\begin{figure}
\caption{ The angular spectra of $\eta$-mesons in the photon-deuteron
c.m. system at different photon energies $E_\gamma$. The data 
are from Ref.~\protect\cite{Hejny}. The dotted lines show
 IA calculations, while the dashed lines represent  the results using only the $np$ final
state interaction. The solid lines show the full calculation, including the
$\eta{N}$ final state interaction from the J\"ulich meson-baryon model.} 
\label{mainz5a}
\end{figure}

\begin{figure}
\caption{ The momentum spectra of $\eta$-mesons in the photon-deuteron
c.m. system at different photon energies $E_\gamma$. The data 
are from Ref.~\protect\cite{Hejny}. The lines show our calculations with
notations similar to Fig.\ref{mainz5a}. The arrows indicates the kinematical
limit for $\eta$-meson momenta calculated by Eq.\ref{max}.} 
\label{mainz5b}
\end{figure}

\begin{figure}
\caption{ The angular (upper part) and momentum (lower part) spectra of 
$\eta$-mesons in the photon-deuteron c.m. system at photon energies 
$E_\gamma$=630-640 and 640-650~MeV. The data 
are from Ref.~\protect\cite{Hejny}. The lines show our calculations 
with different $\eta{N}$ scattering lengths, namely $a_{\eta N}$=0.42+i0.32 
(solid), 0.74+i0.27 (dashed) and 0.25+i0.16~fm (dotted). The arrows 
indicate the kinematical limit for $\eta$-meson momenta calculated 
by Eq.\ref{max}.} 
\label{mainz5u}
\end{figure}

\begin{figure}
\caption{ The invariant mass spectra of the $\eta{p}$ (left)
and $np$ (right) subsystem produced in 
$\gamma{d}{\to}\eta{np}$ reaction at photon energies
$E_\gamma$=643 and 681~MeV. The hatched areas show 
calculations based on the IA, the dashed lines stands for the
calculation including the  $np$ FSI only, while
the solid lines represents the full calculations including
both, $np$ and $\eta N$ FSI.} 
\label{invar1}
\end{figure}



\newpage

\begin{figure}
\begin{center}
\psfig{file=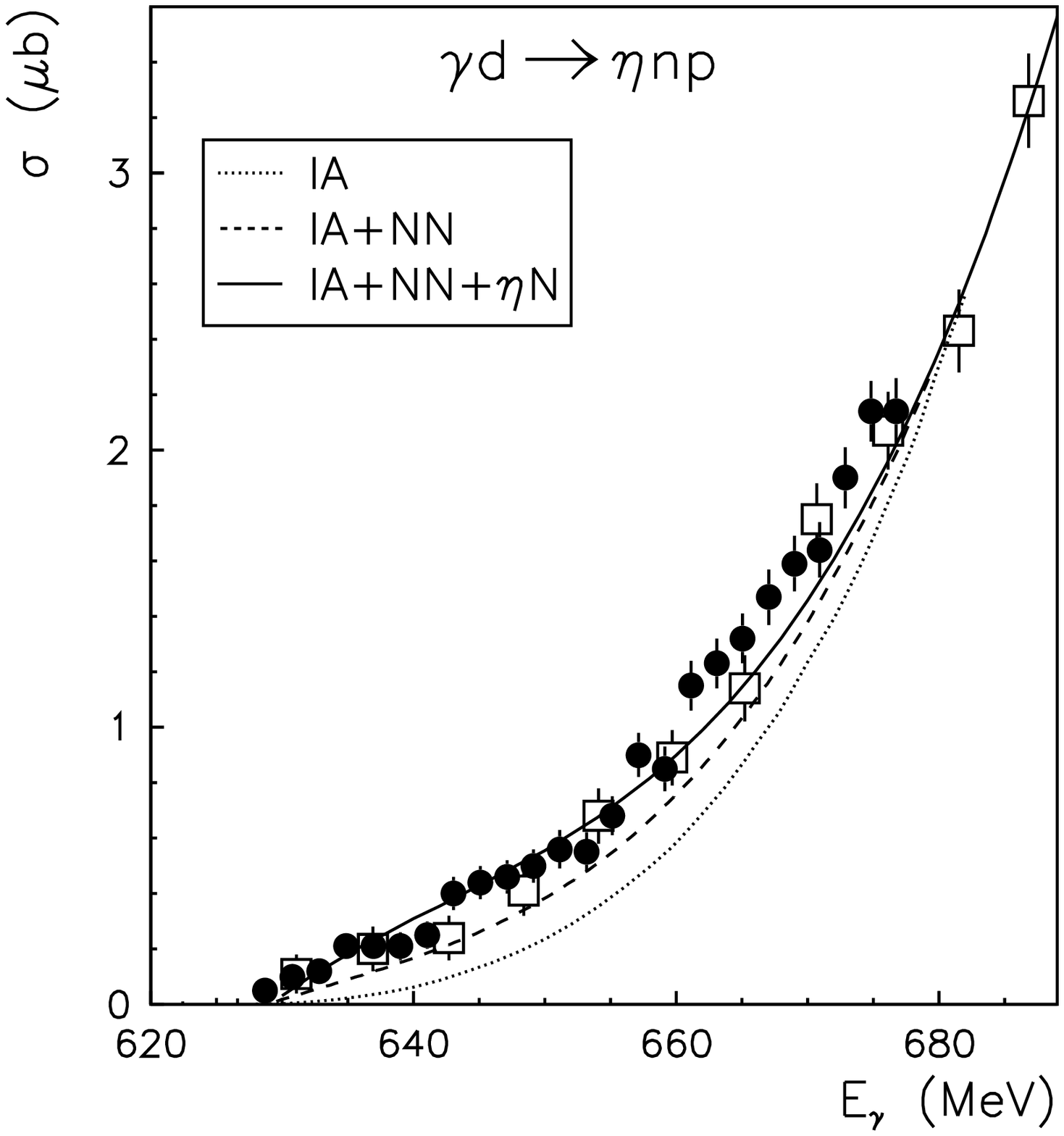,width=10.0cm,height=9.0cm}
\vspace{2mm}
\center{FIG. 1}
\end{center}
\end{figure}

\vspace{-5mm}

\begin{figure}
\begin{center}
\psfig{file=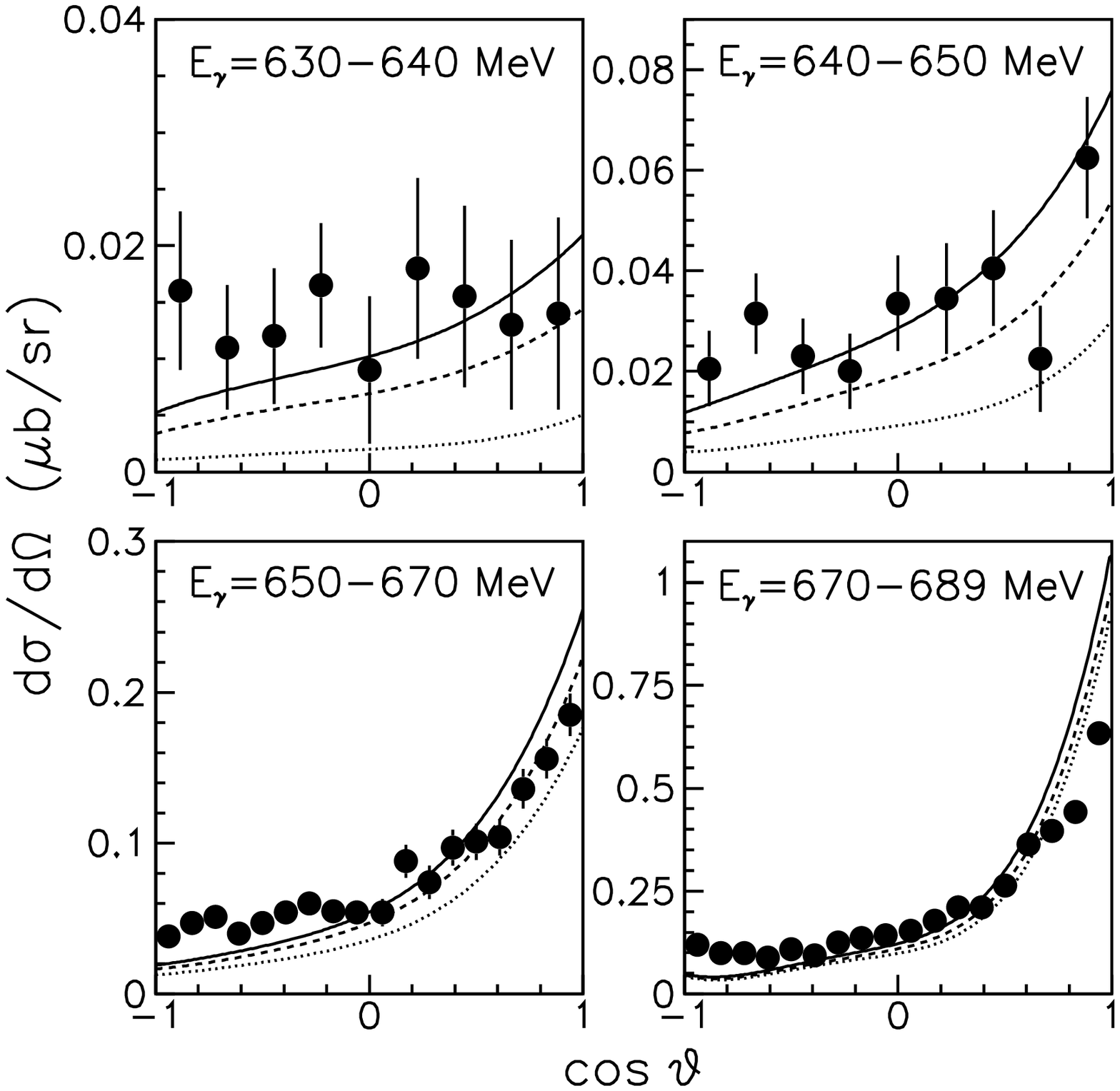,width=10.5cm,height=11.cm}
\vspace{2mm}
\center{FIG. 2}
\end{center}
\end{figure}


\begin{figure}
\begin{center}
\psfig{file=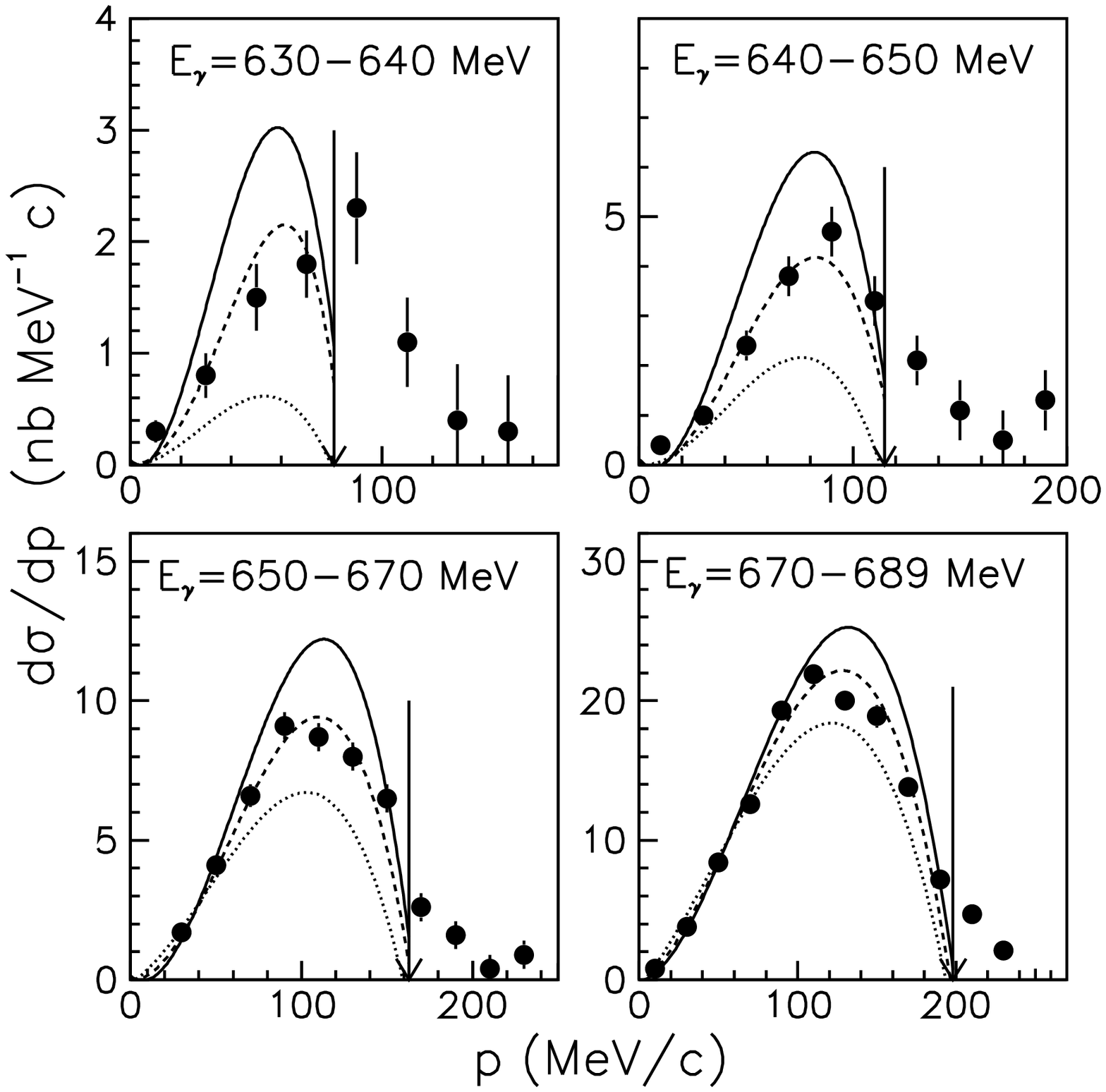,width=10.5cm,height=11.cm}
\vspace{2mm}
\center{FIG. 3}
\end{center}
\end{figure}


\begin{figure}
\begin{center}
\psfig{file=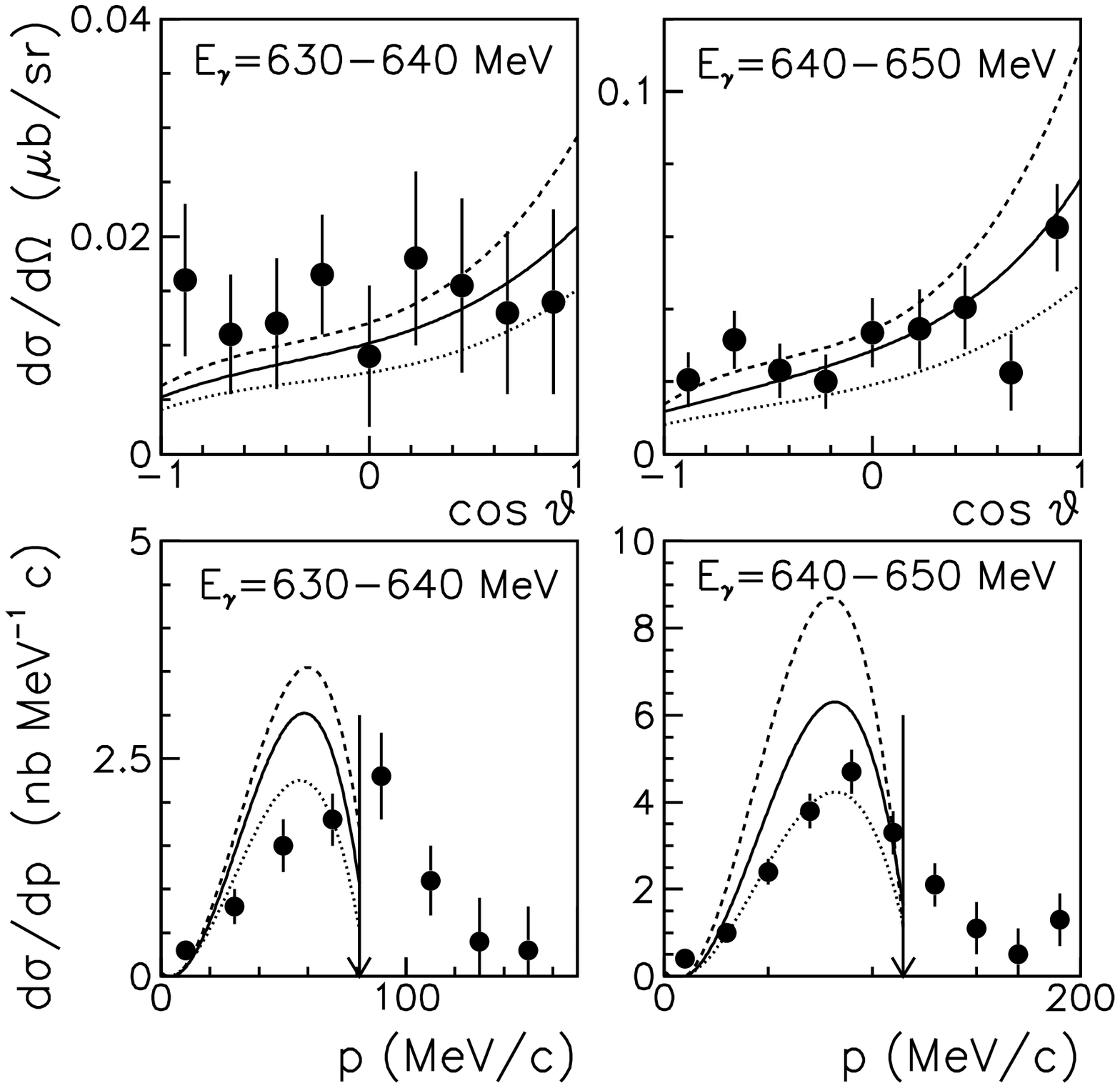,width=10.5cm,height=11.cm}
\vspace{2mm}
\center{FIG. 4}
\end{center}
\end{figure}

\begin{figure}
\begin{center}
\psfig{file=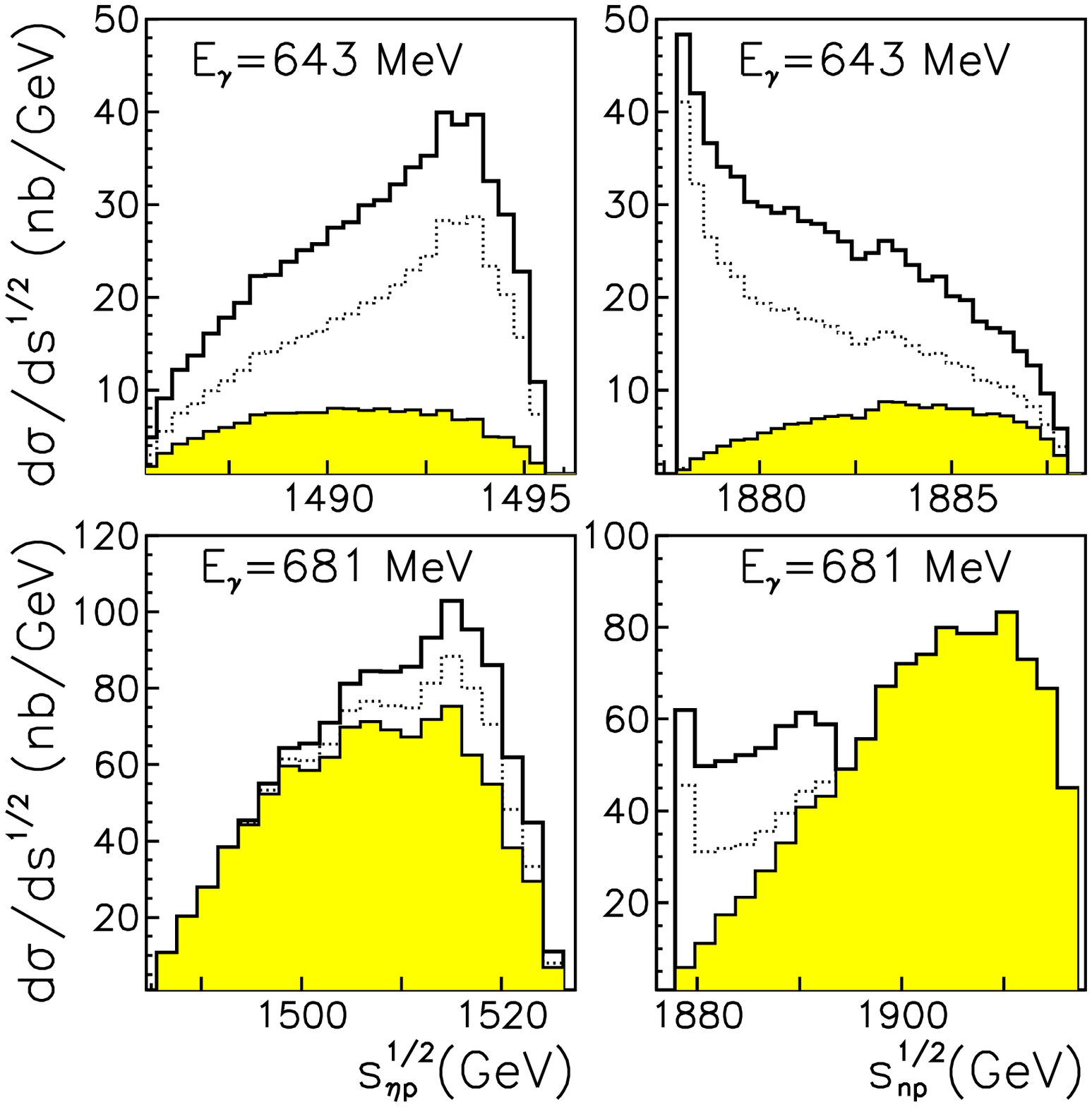,width=10.5cm,height=11.cm}
\vspace{2mm}
\center{FIG. 5}
\end{center}
\end{figure}


\begin{references}
\bibitem{Our1}
        A. Sibirtsev, Ch. Elster, J. Haidenbauer and J.~Speth,
        Phys. Rev. C {\bf 64}, 024006 (2001).
\bibitem{Our2}
        A. Sibirtsev, S. Schneider, Ch. Elster, 
        J. Haidenbauer, S. Krewald and J. Speth,
        to appear in Phys. Rev. C
\bibitem{Krusche1}
        B. Krusche et al., Phys. Lett. B \textbf{358}, 40 (1995).
\bibitem{Hejny}
        V. Hejny et al., to appear in Eur. Phys. J. A. 
\bibitem{Machleidt1} 
        R. Machleidt, Phys. Rev. C \textbf{63}, 024001 (2001).
\bibitem{Delborgo}
        R. Delborgo, Nucl. Phys. \textbf{38}, 249 (1962).
\bibitem{Gillespie}
        J. Gillespie, Final State Interactions, Holden-Day (1964) 91.
\bibitem{Krehl}
        O. Krehl, C. Hanhart, S. Krewald, J. Speth, Phys. Rev. {\bf C62},
        025207 (2000).
\bibitem{Private}
        V. Hejny and H. Str\"oher, private communication.
\bibitem{TOF}
        E. Roderburg and the TOF Collaboration, IKP/COSY Annual Report 2001.
\end{references}
\end{document}